\documentclass[letterpaper, 10 pt, conference]{ieeeconf}  

\IEEEoverridecommandlockouts                              
\usepackage{graphicx} 
\usepackage{amsmath} 
\usepackage{amssymb}  
\usepackage{mathtools} 
\usepackage{siunitx}
\usepackage{wasysym}
\usepackage{float}
\usepackage{xcolor}
\usepackage{hyperref}
\usepackage{xcolor}

\usepackage{amsfonts}
\usepackage{booktabs}
\usepackage{siunitx}

\title{\LARGE \bf
Precipitation event-based networks: an analysis of the relations between network metrics and meteorological properties}

\author{Aurelienne A. S. Jorge$^{1}$, Douglas Uba$^{1}$, Alex A. Fernandes$^{1}$, Izabelly C. Costa$^{1}$, Leonardo B. L. Santos$^{1,3}$ 
\thanks{$^{1}$National Institute for Space Research, Cachoeira Paulista-SP, Brazil.} 
\thanks{$^{2}$Center for Monitoring and Early Warning of Natural Disasters (CEMADEN), S\~{a}o Jos\'{e} dos Campos-SP, Brazil.}  
}

\begin{document}

\maketitle

\begin{abstract}
The study of complex systems in nature is essential to understand the interactions between different elements and how they influence one another. Complex network theory is a powerful tool that helps us to analyze these interactions and gain insights into the behavior of such systems. Surprisingly, this theory has been underutilized in the field of weather science, which focuses on the immediate state of the atmosphere. Our research aims to fill this gap by exploring the use of complex network theory in weather science. Specifically, we employ weather radar data to construct event-based geographical networks. By analyzing the relations between meteorological properties and network metrics in these event-based networks, we can gain a better understanding of the behavior of precipitation events. Our findings reveal significant correlations between various meteorological properties and network metrics, shedding light on the underlying mechanisms that govern precipitation events. Through our work, we hope to demonstrate the potential of complex network theory in weather science and inspire further research in this field.

\end{abstract}

\section{INTRODUCTION}

Complex networks are a powerful tool for analyzing complex systems, such as the climate, using atmosphere remote sensing data. Researchers in the field have employed this approach to identify teleconnection patterns and analyze the structure of climatic events using atmospheric remote sensing data  \cite{Boers2019, SINGHAL2023106538}. To tackle the climate domain, they have employed long time series of atmospheric datasets. The weather, otherwise, is related to short-term changes in the atmosphere, dealing with variables in high resolution both spatially and temporally. 

As pointed out by Akbar $\&$ Saricha (2021) \cite{Akbar2021}, very few works have explored meteorological events in network science. Ceron {\it et al.} (2019) \cite{Ceron2019} is one of the few studies within this context handling high-resolution data from weather radar. With a dataset of only ten days and no description of meteorological events, the author's main finding was that community structures were compatible with the land cover.

We build geographical networks based on precipitation events using weather radar data in the present work. These events occurred from January to March 2019, focusing on the Metropolitan Area of São Paulo (MASP). Based on this set of events, we analyze the relations between the meteorological properties and the topological metrics of the correspondent networks. Our findings show significant correlations and some particularities when analyzing specific event groups.
\section{MATERIAL AND METHODS}

\subsection{Data}

For our study case, we used data from a weather radar in the city of São Roque, whose coverage includes the entire study area (described in the next section). Its operation is in charge of the Department of Airspace Control (DECEA), and its range is 250 km in qualitative mode and 400 km in surveillance mode. This weather radar performs a volumetric scan of azimuth scans in 15 different elevation angles, from 0.5 to nearly 20 degrees. 
We can produce diverse products from such volumetric scans, including CAPPI (Constant Altitude Position Indicator) that is a projection of a horizontal plane at a constant height. In this work, we use the CAPPI product at the height of 3 km, which avoids altitude changes and ground echo problems. The values are used in reflectivity units (dBZ) as the product supplies them.

Using such data, we analyze precipitation events from January to March 2019. We used a computational tool named TATHU (Tracking and Analysis of Thunderstorms) \cite{tathu} to identify spatial and temporal events for automatically tracking weather systems' life cycles. Applying an overlap verification between the time steps, the tool can associate the features in a timeline, delineating the weather event. 
The parameters we define for this work are: a minimum area of 9 squared kilometers and values equal to or above 20 dBz. Such a reflectivity threshold intends to avoid noise signals and filter only values equivalent to 1 millimeter per hour or above. An overlap of at least $10\%$ of the area is required to track the event.

Our study area in this work is the metropolitan area of São Paulo (MASP). Brazil's most significant metropolitan region has 39 municipalities and more than 19 million inhabitants. From the events database we produced, we select those which occur inside the MASP bounding box, adding a buffer of 10 km. Next, we filter the events whose duration is at least $1$ hour and $40$ minutes, and at most $20$ hours. The lower threshold guarantees at least $10$ time steps to calculate correlations later. The upper threshold is to avoid huge events with a high computational cost. After applying these filters, we end up with a sample of $383$ events. We build a geographical network for each, selecting the weather radar dataset according to the event's duration. 

\subsection{Precipitation event networks}
\label{sec:event_nets}

In the next step, such dataset is used as input to \href{http://github.com/aurelienne/graph4gis}{Graph4GIS} \footnote{http://github.com/aurelienne/graph4gis}, an application developed in Python able to build geographical networks from gridded binary data. As a first step, G4G converts each grid point of the dataset, inside the selected study area, into a network node carrying an attribute of geographical coordinate. Then, the software selects the time series associated with every grid point and binds it to the corresponding node. Nodes with a completely zeroed time series are discarded from the network. 

The next step is to calculate the similarities between the pair of nodes. We adopt Pearson Correlation as the similarity function for the present case study, adding an option of time delay in the calculations. This delay ranges from $0$ to $30$ minutes, and we keep the delay that maximizes the correlation for each pair of nodes. In the end, we have two filled matrices: one with the delays and the other with the respective correlations (weight matrix). 

The Global Threshold (GT) criterion is applied to the weight matrix to select the most relevant weights converted into edges in our graph. For this research, we define the GT value as the point of maximum diameter of the network. This way, we intend to promote the best possible balance between removing the least relevant edges and keeping the most important ones - as applied in previous papers in the literature \cite{Ceron2019}. At the end of G4G processing, we have a geographical network for the precipitation event with associated network metrics. Besides the number of links/edges (L) and nodes (N), the following metrics are included for analysis: average of the local clustering coefficients ($\langle c \rangle$), average degree ($\langle k \rangle$), diameter ($D$), average shortest path ($\langle l \rangle$), number of components ($NC$), giant component ($GC$) and singletons ($ST$). The local clustering coefficient indicates how connected a node's neighbors are to each other. $\langle k \rangle$ is the average number of connections in the network. The diameter is defined as the longest shortest path of a network. $\langle l \rangle$ refers to the average shortest paths linking every pair of vertices. $NC$ is the number of isolated groups of nodes (components), $GC$ is the size of the largest component, and $ST$ is the number of components with a single node.

From the database of selected events, we extract a set of physical properties that characterize each of them, in terms of area, duration and intensity. We call these properties meteorological ("meteo") metrics. G4G builds the correspondent network for each event and calculates its topological metrics. A Person Correlation is performed for every pair meteo-network metric, discarding those correlations with a $p-value$ not statistically significant ($alpha=0.05$). The result is a graph representing these relations between network and meteorological metrics.

\section{RESULTS AND DISCUSSION}

\begin{figure}[h!]
\centering 
\caption{Meteo-Network Graph: Network metrics (orange nodes), Meteorological properties (grey nodes), positive and negative correlations (blue and red edges, respectively). }
 \makebox[\columnwidth]{\includegraphics[width=\linewidth]{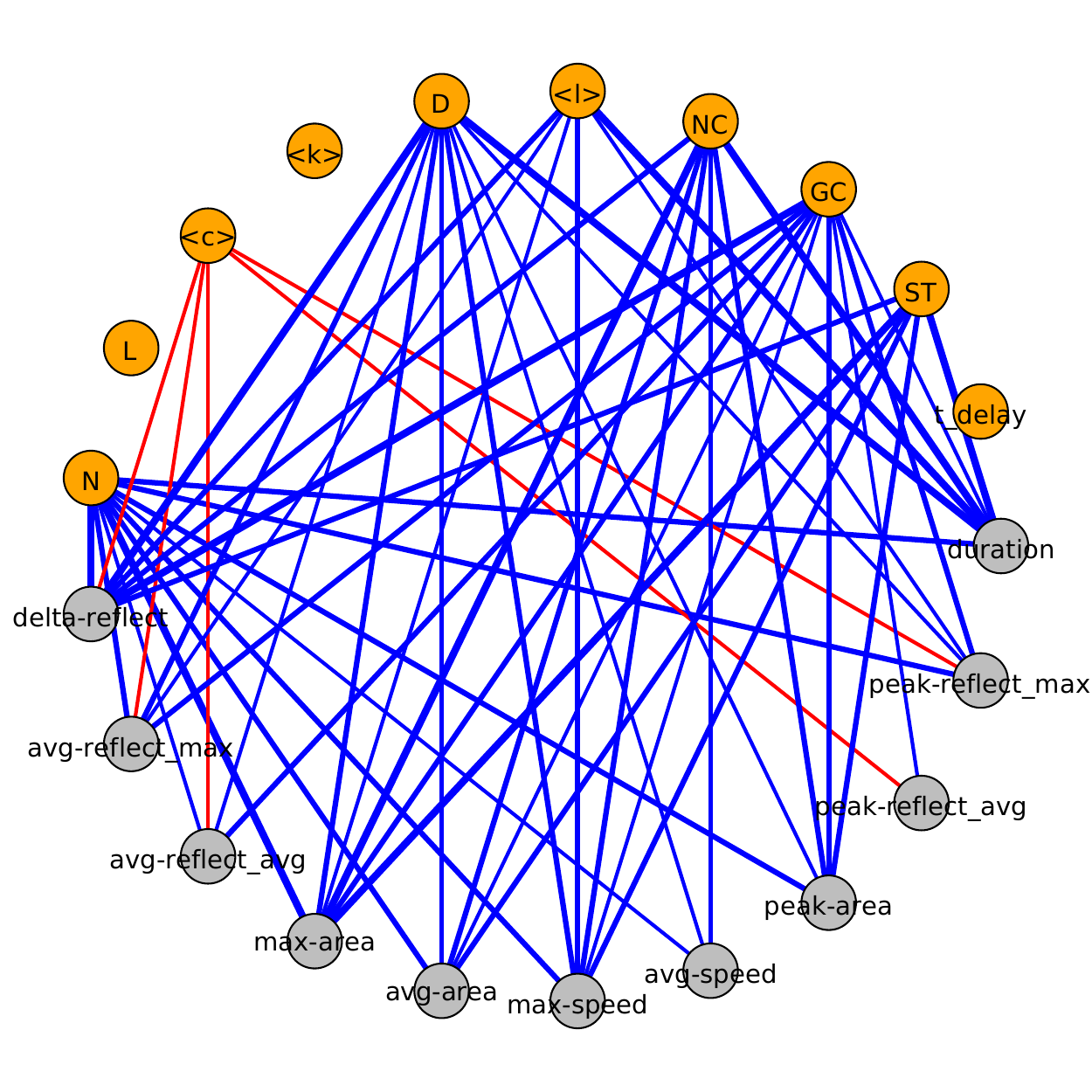}}
\label{general_graph}
\end{figure}

Figure \ref{general_graph} shows the resulting graph with the meteo-network correlations for the entire set of selected events (sample with $383$ events as mentioned in \ref{sec:event_nets}). The orange nodes are the network metrics, the grey nodes are the meteorological properties, the blue edges represent positive correlations, and the red edges are negative correlations. The thicker the edges, the higher the correlation modulus. Only correlation coefficients above $0.4$ or below $-0.4$ are included in the graph. Every edge connects a network node to a meteorological node, resulting in a bipartite graph. 

$N$ (number of nodes), $GC$ (giant component), and $D$ (diameter) are the nodes with the most significant number of connections, each one with $10$. As the event has a larger area, greater duration, or higher speed, the correspondent network tends to spread to follow the event track. Consequently, it results in a higher number of vertices. The result shows that the number of components also tends to increase as well as the size of the giant component. The paths also tend to expand with a more extensive network as the clustering coefficient does not correlate positively. Therefore, the diameter metric increases with the event's area, speed, or duration. 

We can also notice that the higher the event's reflectivity variation ("delta-reflect") is, the greater the network is, as there is a positive correlation with the $N$ property. It shows that events with a broader reflectivity range tend to be those with larger areas or greater duration. For the same reason, this reflectivity variation also positively correlates with the diameter and the average shortest path. Differently, the clustering coefficient has a negative correlation with the reflectivity measures. The higher the reflectivity values are, the less clustered the network is. Higher values in reflectivity time series probably become more challenging to have similarities, affecting the network clustering.

However, the mentioned results are derived from the entire set of events, including various meteorological processes. Intending to analyze these correlations in more specific scenarios, we define two group categories by classifying events by duration. The first one, named $D1$, includes short-duration events ($2$ hours or less), which may be related to local convective systems that generally present high intensity in a very brief occurrence. The second group, named $D2$, contains long-duration events ($5$ hours or more), which may be associated with weather fronts - meteorological processes that usually affect larger areas within greater periods. There are $114$ events in $D1$, whereas $D2$ has $53$ events. Our dataset comprises the summer period, so convective systems are naturally expected since air humidity is higher this year's season.

The resulting graph concerning the D1 group presents some particularities compared with the general graph. One is the positive correlation between the average degree and the event's average reflectivity ("avg-reflect\_avg"). In short events, the average reflectivity increases homogeneously, supporting more connections. The positive correlations involving area, speed, or duration do not appear for brief events. In the context of extended events ($D2$ group), the clustering coefficient presents a negative correlation with duration, area ("avg", "max" and "peak"), and "max-speed", which we do not see in the general graph. The diameter and average shortest paths do not correlate with the area size, as the network paths are more related to the duration in this scenario. The positive correlation between duration and $D/\langle l \rangle$ corroborates that. On the other hand, the number of edges positively correlates with maximum and average areas. Similarly, the number of connected components, singletons, and the giant components' size correlate to the area dimension.

\begin{figure}[h!]
\centering 
\caption{Boxplots of network metrics: Diameter (D), Number of Components (NC), and Average Shortest Path ($\langle l \rangle$), from D1 (blue) and D2 (red) groups.}
 \makebox[\columnwidth]{\includegraphics[width=\linewidth]{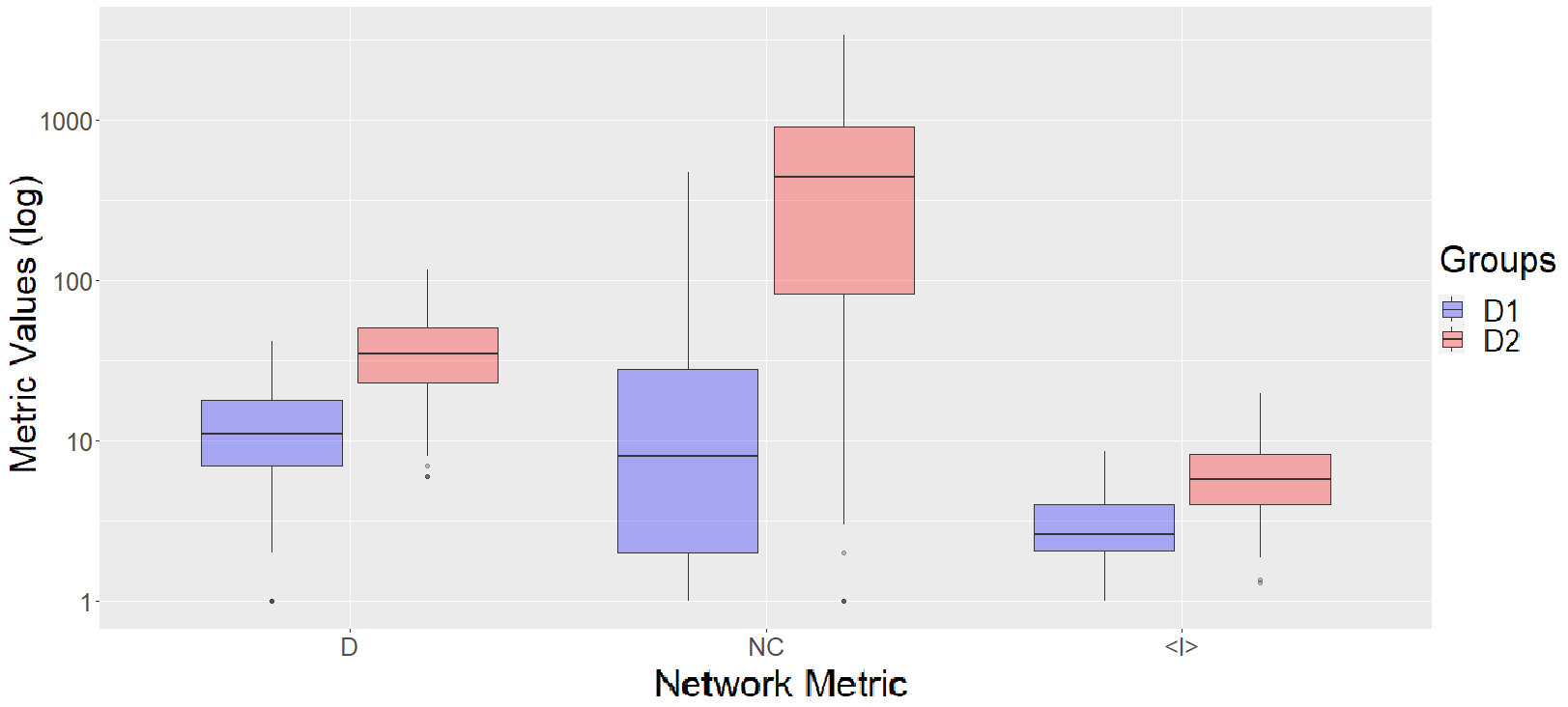}}
\label{boxplots}
\end{figure}

In relation to the groups, we also compare the sample sizes of each group/metric analyzed. Figure \ref{boxplots} presents the boxplots of the following samples: $NC$, $D$ and $\langle l \rangle$ from $D1$ and $D2$ groups. It is possible to verify that the samples are greater for the group of long-duration events. As previously discussed, it is naturally expected to have a wider network as the event duration extends. And the results confirm that behavior with significant and positive correlation values between the event duration and metrics such as the diameter, number of components, and the average shortest path. We applied the Mann-Whitney test to verify if the distributions underlying $D2$ samples exceed $D1$. It is a non-parametric test to test the null hypothesis that the distributions underlying two samples are equal. The null hypothesis was rejected for all the comparisons with a significant $p-value$ ($alpha=0.05$), pointing out that the distributions underlying all $D2$ samples are greater than $D1$ samples. 

\section{CONCLUSION}

This work presented the analysis of network metrics applied in the weather scope based on a set of precipitation events. We examined the relations between meteorological properties and network metrics, and it was done in two scenarios: the entire set of events and groups of events based on their duration.

Concerning the general ensemble, we could see that a larger area, greater duration, or higher speed tends to extend the correspondent network following the event track. Consequently, metrics such as the number of components and the diameter also tend to increase. The clustering coefficient, otherwise, decreases as the reflectivity values vary. We also analyzed the meteo-network relations inside specific groups of events, classifying them by duration. It was possible to notice some particularities in these scenarios. In short-duration events, the average reflectivity increases homogeneously, supporting more connections and increasing the network's average degree. Concerning long-duration events, there is a negative correlation between the clustering coefficient and the event's area or duration. In other words, the more extended the event is, the less clustered its network is. 

Continuing this work, we plan to analyze the projections of the network and meteorological sets in the bipartite graphs, aiming to go deeper into the correlation analysis. We also intend to embrace other periods in future works and combine different atmospheric variables in multi-layer networks, including forecast data. The expectation is to anticipate extreme events in a very short term.
\section*{ACKNOWLEDGMENT}
{\footnotesize
This work was supported by FAPESP Grant Numbers 2015/50122-0, 2018/06205-7; DFG-IRTG Grant Number 1740/2.
}


\footnotesize
\bibliography{sec/refs}
\bibliographystyle{IEEEtran}

\end{document}